\documentclass[fleqn,twoside]{article}
\usepackage{espcrc2}
\usepackage{axodraw}

\newcommand{\la}[1]{\label{#1}}
\newcommand{\be}{\begin{equation}}
\newcommand{\ee}{\end{equation}}
\newcommand{\ba}{\begin{eqnarray}}
\newcommand{\ea}{\end{eqnarray}}

\newcommand{\eq}{Eq.~}
\newcommand{\nr}[1]{(\ref{#1})}
\newcommand{\nn}{\nonumber \\}
\newcommand{\fr}[2]{{\frac{#1}{#2}\,}}
\renewcommand{\(}{\left(}
\renewcommand{\)}{\right)}

\newcommand{\lk}{\left[}
\newcommand{\rk}{\right]}
\newcommand{\ld}{\left.}

\newcommand{\e}{\epsilon}

\newcommand{\rmi}[1]{{\mbox{\scriptsize #1}}}
\newcommand{\aG}{\alpha_\rmi{G}}
\newcommand{\bG}{\beta_\rmi{G}}
\def\Asc(#1,#2)(#3,#4,#5){\CArc(#1,#2)(#3,#4,#5)}
\def\Lsc(#1,#2)(#3,#4){\Line(#1,#2)(#3,#4)}
\def\Ahh(#1,#2)(#3,#4,#5){\DashCArc(#1,#2)(#3,#4,#5){1}}
\def\Lhh(#1,#2)(#3,#4){\DashLine(#1,#2)(#3,#4){1}}
\def\Aqu(#1,#2)(#3,#4,#5){\ArrowArc(#1,#2)(#3,#4,#5)}
\def\Aaqu(#1,#2)(#3,#4,#5){\ArrowArcn(#1,#2)(#3,#5,#4)}
\def\scfc{0.7}  
\def\phgt{21}   
\def\pwc{21}    
\def\pwcb{31.5} 
\newcommand{\PIC}[4]{\;\parbox[c]{#2 pt}{\begin{picture}(#2,#3)(0,0)
\SetWidth{1.0}\SetScale{#4} #1 \end{picture}}\;}
\newcommand{\pic}[1]{\PIC{#1}{\pwc}{\phgt}{\scfc}}
\newcommand{\picb}[1]{\PIC{#1}{\pwcb}{\phgt}{\scfc}}
\def\TopoVR(#1){\pic{#1(15,15)(15,-90,270)}}
\def\ToptVS(#1,#2,#3){\pic{#1(15,15)(15,0,180) #2(15,15)(15,180,360)%
 #3(30,15)(0,15)}}
\def\ToprVM(#1,#2,#3,#4,#5,#6){\pic{#3(15,15)(15,-30,90) #1(15,15)(15,90,210)%
 #2(15,15)(15,210,330) #5(2,7.5)(15,15) #6(15,15)(15,30) #4(28,7.5)(15,15)}}
\def\ToprVMsp(#1,#2,#3,#4,#5,#6){\pic{#3(15,15)(15,-30,90)%
 #1(15,15)(15,90,210)#2(15,15)(15,210,330) #5(2,7.5)(15,15)%
 #6(15,15)(15,30) #4(28,7.5)(15,15)\Text(19,18)[bl]{$\scriptstyle 2m$}}}
\def\ToprVV(#1,#2,#3,#4,#5){\!\!\picb{#2(26.25,15)(15,256,76)%
 #3(30,30)(15,30) #1(18.75,15)(15,104,284) #4(15,30)(22.5,0)%
 #5(30,30)(22.5,0)}\!\!}
\def\ToprVB(#1,#2,#3,#4){\picb{#1(30,15)(15,-120,120) #2(30,15)(15,120,240)%
 #3(15,15)(15,60,300) #4(15,15)(15,-60,60)}}
\def\TopfVX(#1,#2,#3,#4,#5,#6,#7,#8,#9){\picb{#1(15,15)(15,90,270)%
 #2(30,15)(15,-90,90) #4(30,30)(15,30) #3(15,0)(30,0) #6(15,0)(15,15)%
 #5(15,15)(30,30) #8(15,30)(20,25) #8(25,20)(30,15) #7(30,15)(30,0)%
 #9(15,15)(30,15)}}
\def\TopfVH(#1,#2,#3,#4,#5,#6,#7,#8,#9){\picb{#1(15,15)(15,90,270)%
 #2(30,15)(15,-90,90) #4(30,30)(15,30) #3(15,0)(30,0) #6(15,0)(15,15)%
 #5(15,15)(15,30) #8(30,30)(30,15) #7(30,15)(30,0) #9(15,15)(30,15)}}
\def\TopfVW(#1,#2,#3,#4,#5,#6,#7,#8){\pic{#1(15,15)(15,90,180)%
 #3(15,15)(15,180,270) #2(15,15)(15,270,360) #4(15,15)(15,0,90)%
 #5(15,15)(15,30) #7(15,15)(15,0) #6(0,15)(15,15) #8(30,15)(15,15)}}
\def\TopfVWdot(#1,#2,#3,#4,#5,#6,#7,#8){\pic{#1(15,15)(15,90,180)%
 #3(15,15)(15,180,270) #2(15,15)(15,270,360) #4(15,15)(15,0,90)%
 #5(15,15)(15,30) #7(15,15)(15,0) #6(0,15)(15,15) #8(30,15)(15,15)%
 \Vertex(22,15){2}}}
\def\TopfVWlap(#1,#2,#3,#4,#5,#6,#7,#8){\pic{#1(15,15)(15,90,180)%
 #3(15,15)(15,180,270) #2(15,15)(15,270,360) #4(15,15)(15,0,90)%
 #5(15,15)(15,30) #7(15,15)(15,0) #6(0,15)(15,15) #8(30,15)(15,15)%
 \Text(2,19)[br]{$\scriptstyle 1$}\Text(20,2)[tl]{$\scriptstyle 2$}}}
\def\TopfVV(#1,#2,#3,#4,#5,#6,#7,#8){\!\!\picb{#2(26.25,15)(15,256,346)%
 #3(26.25,15)(15,-14,76) #4(30,30)(15,30) #1(18.75,15)(15,104,284)%
 #7(22.5,0)(15,30) #6(30,30)(26.25,15) #8(26.25,15)(22.5,0)%
 #5(25.25,15)(39.8,11.4)}\!\!}
\def\TopfVB(#1,#2,#3,#4,#5,#6,#7){\picb{#2(30,15)(15,-120,120)%
 #6(30,15)(15,120,180) #5(30,15)(15,180,240) #1(15,15)(15,60,300)%
 #4(15,15)(15,-60,0) #3(15,15)(15,0,60) #7(30,15)(15,15)}}
\def\TopfVBdot(#1,#2,#3,#4,#5,#6,#7){\picb{#2(30,15)(15,-120,120)%
 #6(30,15)(15,120,180) #5(30,15)(15,180,240) #1(15,15)(15,60,300)%
 #4(15,15)(15,-60,0) #3(15,15)(15,0,60) #7(30,15)(15,15)%
 \Vertex(28,22){2}}}
\def\TopfVBdotNew(#1,#2,#3,#4,#5,#6,#7){\picb{#2(30,15)(15,-120,120)%
 #6(30,15)(15,120,180) #5(30,15)(15,180,240) #1(15,15)(15,60,300)%
 #4(15,15)(15,-60,0) #3(15,15)(15,0,60) #7(30,15)(15,15)%
 \Vertex(45,15){2}}}
\def\TopfVBlap(#1,#2,#3,#4,#5,#6,#7){\picb{#2(30,15)(15,-120,120)%
 #6(30,15)(15,120,180) #5(30,15)(15,180,240) #1(15,15)(15,60,300)%
 #4(15,15)(15,-60,0) #3(15,15)(15,0,60) #7(30,15)(15,15)%
 \Text(9,15)[r]{$\scriptstyle 1$}\Text(22.5,15)[l]{$\scriptstyle 2$}}}
\def\TopfVN(#1,#2,#3,#4,#5,#6,#7){\picb{#1(15,15)(15,90,270)%
 #2(30,15)(15,-90,90) #4(30,30)(15,30) #3(15,0)(30,0)%
 #5(15,0)(15,30) #6(30,30)(30,0) #7(15,30)(30,0)}} 
\def\TopfVU(#1,#2,#3,#4,#5,#6,#7){\pic{#3(15,15)(15,0,90)%
 #2(15,15)(15,90,180) #4(15,15)(15,180,270) #1(15,15)(15,270,360)%
 #6(0,15)(15,30) #7(15,0)(0,15) #5(30,15)(15,0)}}
\def\TopfVT(#1,#2,#3,#4,#5,#6){\pic{#1(15,15)(15,90,210)%
 #2(15,15)(15,210,330) #3(15,15)(15,-30,90) #4(2,7.5)(15,30)%
 #6(28,7.5)(2,7.5) #5(15,30)(28,7.5)}}
\def\TopLV(#1,#2,#3,#4,#5,#6){\!\!\picb{#2(26.25,15)(15.5,256,76)%
 #3(30,30)(15,30) #1(18.75,15)(15.5,104,284) #4(15,30)(22.5,0)%
 #5(30,30)(22.5,0) #6(15,17.8)(19.3,292.8,39.1)}\!\!}
\def\TopLVdot(#1,#2,#3,#4,#5,#6){\!\!\picb{#2(26.25,15)(15.5,256,76)%
 #3(30,30)(15,30) #1(18.75,15)(15.5,104,284) #4(15,30)(22.5,0)%
 #5(30,30)(22.5,0) #6(15,17.8)(19.3,292.8,39.1)\Vertex(26.25,15){2} }\!\!}
\def\TopfVBB(#1,#2,#3,#4,#5){\picb{#1(30,15)(15,-120,120)%
 #2(30,15)(15,120,240) #3(15,15)(15,60,300) #4(15,15)(15,-60,60)%
 #5(22.5,3)(22.5,27)}}
\def\TopfVBBdot(#1,#2,#3,#4,#5){\picb{#1(30,15)(15,-120,120)%
 #2(30,15)(15,120,240) #3(15,15)(15,60,300) #4(15,15)(15,-60,60)%
 #5(22.5,3)(22.5,27) \Vertex(22.5,10){2} \Vertex(22.5,20){2}}}
\def\TopfVBBlap(#1,#2,#3,#4,#5){\picb{#1(30,15)(15,-120,120)%
 #2(30,15)(15,120,240) #3(15,15)(15,60,300) #4(15,15)(15,-60,60)%
 #5(22.5,3)(22.5,27)%
 \Text(2,10.5)[l]{$\scriptstyle 1$}\Text(29.5,10.5)[r]{$\scriptstyle 2$}}}

\def\one{\TopoVR(\Asc)}
\def\two{\ToptVS(\Asc,\Asc,\Lsc)}

\def\threeMspecial{\ToprVMsp(\Asc,\Asc,\Asc,\Lsc,\Lsc,\Lsc)}

\def\topoII{\TopfVW(\Asc,\Asc,\Asc,\Asc,\Lsc,\Lsc,\Lsc,\Lsc)} 

\def\topoIV{\TopfVB(\Asc,\Asc,\Asc,\Asc,\Asc,\Asc,\Lsc)}
\def\topoIVxtraNew{\TopfVBdotNew(\Asc,\Asc,\Asc,\Asc,\Asc,\Asc,\Lsc)}

\def\topoVIII{\TopfVT(\Asc,\Asc,\Asc,\Lsc,\Lsc,\Lsc)} 
\def\topoIX{\TopLV(\Asc,\Asc,\Lsc,\Lsc,\Lsc,\Asc)} 
\def\topoIXxtra{\TopLVdot(\Asc,\Asc,\Lsc,\Lsc,\Lsc,\Asc)} 
\def\topoXII{\TopfVBB(\Asc,\Asc,\Asc,\Asc,\Lsc)} 

\def\topoIIxtra{\TopfVWdot(\Asc,\Asc,\Asc,\Asc,\Lsc,\Lsc,\Lsc,\Lsc)}
\def\topoIVxtra{\TopfVBdot(\Asc,\Asc,\Asc,\Asc,\Asc,\Asc,\Lsc)}
\def\topoXIIxtra{\TopfVBBdot(\Asc,\Asc,\Asc,\Asc,\Lsc)} 

\newcommand{\us}[1]{\frac{1}{d-x}}

\newcommand{\bmu}{\bar\mu}

\title{Tackling the infrared problem of 
thermal QCD\hfill {\normalsize MIT-CTP-3421}}
\author{
Y.~Schr\"{o}der \address[CTP]{Center for Theoretical Physics, MIT, 
Cambridge, MA 02139, USA}%
\thanks{talk presented at Lattice 2003.}
}
\begin{document}


\begin{abstract}
Perturbative calculations of corrections to the behavior 
of an ideal gas of quarks and gluons, 
the limit that is formally realized at infinite temperature,
are obstructed by severe infrared divergences.
The limits to computability that the infrared 
problem poses can be overcome in the framework of dimensionally 
reduced effective theories. 
Here, we give details on the evaluation of the highest
perturbative coefficient needed for this setup, in the continuum.
\vspace{-3mm}
\end{abstract}


\maketitle


\section{INTRODUCTION}

The theory of strong interactions, Quantum Chromodynamics (QCD),
is guaranteed to be accessible to perturbative methods once
one of its parameters, the temperature $T$, is increased towards
asymptotically high values.
This general statement relies solely on the well-known property of
asymptotic freedom.
 
In practice, however, calculations of corrections to the behavior
of an ideal gas of quarks and gluons,
the limit that is formally realized at infinite $T$,
are obstructed by severe infrared (IR) 
divergences \cite{irproblem}: for every observable,
there exists an order of the perturbative
expansion to which an infinite number of Feynman diagrams contribute.
No method is known how to re-sum these infinite classes of diagrams,
a fact that seriously obstructs progress in the field of thermal
QCD. 
 
It is known how to evade this obstruction using 
dimensionally reduced effective theories. 
The key idea is to map the
infrared sector of thermal QCD onto a three-dimensional pure gauge
theory \cite{irproblem,dr,bn,gsixg}, whose contribution, being a pure number, 
could be extracted numerically by Monte-Carlo simulations.
While the expansion of the QCD pressure in the effective theory
framework, up to the order where IR contributions are relevant,
is now known analytically \cite{gsixg},    
realizing the numerical extraction of the yet-unknown number emerging
from the IR sector is a challenging open problem,
with the main complication that high-order matching between 
lattice and continuum regularization schemes is necessary \cite{nspt}.


\section{SETUP}

Let us now switch gears and focus on one of the main
building blocks of the procedure,
while for a detailed description of the setup as well as notation 
and further references, we refer to \cite{gsixg}.
In particular, we want to compute the (negative) 3d vacuum energy density 
of a pure SU(N) gauge theory,
\ba
\lim_{V\to\infty} \fr1V \ln\int{\cal D}A_i\,
\exp\(-\int{\rm d}^dx\,\fr12 {\rm Tr}F_{ij}^2\) \;,
\ea
which in a weak-coupling expansion can be written as the sum of all 
connected vacuum graphs containing gluons and ghosts.
Since the theory is confining, the computation involves IR 
divergent integrals (starting at the 4-loop level here), 
forbidding a perturbative evaluation of the full vacuum energy.
One can however obtain its logarithmic ultraviolet divergence. 

Note that in 3d the coupling constant $g$ is dimensionful, 
hence the full answer must be of the form
\ba
d_A C_A^3 \frac{g^6}{(4\pi)^4} 
\lk \aG \( \frac1\e + 8 \ln\frac{\bmu}{2m_\rmi{M}}\) 
+\!\bG +\!{\cal O}(\e) \rk \;, \nonumber
\ea
where $m_\rmi{M} \equiv C_A g^2$ is a dynamically generated
infrared scale in the confining theory, and $C_A=N$ and $d_A=N^2-1$
are the Casimir and the dimension of the adjoint representation,
respectively.
Because of super-renormalizability, 
the coefficient $\aG$ can then be computed in 4-loop perturbation theory, 
even if the constant part $\bG$ cannot.

If we just carry out the 4-loop
computation in strict dimensional regularisation, the result vanishes, 
because there are no perturbative mass scales in the problem. 
This means that UV and IR divergences
(erroneously) cancel against each other. 
Therefore, we have to be more careful in order to determine $\aG$. 
To regulate the IR divergences, we introduce by hand a mass scale, 
$m^2$, into the gauge field (and ghost) propagators. 
One has to keep in mind, however, that now only the coefficient 
$\aG$ multiplying $1/\epsilon$ is physically meaningful, 
as it contains the desired gauge independent ultraviolet
divergence. 
On the contrary, the constant part  
depends on the gauge parameter $\xi$, because
the introduction of $m^2$ breaks gauge invariance, 
and has nothing to do with $\bG$. 

Note that e.g. diagrams with self-energy insertions
can have IR sub-divergences, since IR divergences are known to be present 
in the 3d 2-loop gluon propagator. 
To avoid the problem of overlapping IR divergences from the outset,
we have hence chosen to employ the mass parameter rigorously, i.e.
by rewriting every $1/p^2$ as $1/(p^2+m^2)$. 

This leaves us within the class of fully massive integrals. 
The computation can be divided in three parts.
Roughly, those are (1) diagram generation \cite{diags}, 
specification of
Feynman rules and color algebra, (2) reduction to master
integrals \cite{reduction,master}, (3) expansion in $d=3-2\e$ dimensions.

We will refrain from commenting on the first two parts of the
computation here, since they are well documented in the references
given above. Due to the complexity of the computation, both steps
are automatized, 
allowing for
the handling of a large set of diagrams.


\section{MASTER INTEGRAL REPRESENTATION}

Let us now give a little more detail on part (3) of the computation.
At this point, all diagrams are expressed in terms of 19 scalar master
integrals,
which are enumerated in \cite{master}.
The general structure is
\ba \label{eq:Sum}
d_A C_A^3 \fr{g^6}{(4\pi)^4} \sum_{i=1}^{19} 
\fr{{\rm poly}_i(d,\xi)}{{\rm poly}_i(d)} {\rm Master}_i(d) \;,
\ea
where $d$ is still an arbitrary (space-time) dimension.
Only now do we need to specify $d=3-2\e$.

While it is trivial to expand the polynomial prefactor in $\e$,
considerable effort has to be put into obtaining the expansion
for the master integrals to the depth required. 
Since we need the $\e$-poles only, it would seem sufficient to
compute the divergent parts of all master integrals. It turns
out, however, that the prefactor develops poles as well around
3 dimensions, having terms proportional to $1/(d-3)$ multiplying 
10 of the master integrals, and even double poles in 4 of those cases.

A crucial simplification can be made by exploiting the freedom
of choosing the basis of master integrals to represent the sum
of diagrams \eq\nr{eq:Sum}. Going back to the tabulated relations
between integrals that were derived by partial integration and 
used in part (2), we found two most useful relations:
\ba
\topoXIIxtra &=& 
-\fr{8(d-3)}5 \topoIXxtra\nn&&
-\fr{(d-3)(3d-8)}5 \topoIX\nn&&
+\fr{(2d-7)(2d-5)}{25} \topoXII \nn&&
-\fr{(d-2)^2}{10} \(\!\!\one\!\!\)^2\!\!\two
\;,\\
\topoIVxtra&=&
-\fr23\topoIVxtraNew
-\fr{3d-10}6\topoIV\nn&&
+\fr13\topoIXxtra
+\fr{d-3}9 \topoVIII
\;.
\ea
Notation: each line represents
a massive scalar propagator, a dot on a line means an extra power,
vertices have no structure.
Trading the two master integrals on the lhs of the above equations
for the first ones on the rhs respectively (all others are already
included in the basis),
the $d$-dimensional representation \eq\nr{eq:Sum}
of course still holds, albeit with a `primed' version
of the basis,
\ba \label{eq:prime}
d_A C_A^3 \fr{g^6}{(4\pi)^4} \sum_{i=1}^{19} 
\fr{{\rm poly}_i^\prime(d,\xi)}{{\rm poly}_i^\prime(d)} {\rm Master}_i^\prime(d) \;.
\ea
In this new basis, none of the prefactors has a double pole
in 3d, while only 7 members of the new `primed' basis are multiplied
by a single pole. 
It is not excluded that there exists a choice of basis for which 
the prefactors never get singular, but this choice is currently not known 
to us.


\section{EXPANSION}

It turns out that (almost) all integrals are known analytically
to the order needed for obtaining the poles in the sum
of all diagrams. Lower loop cases have been treated in \cite{ints3loop},
while analytic results for the divergences of all 3d 4-loop master 
integrals as well as numerical and some analytic results for their
constant parts as well as the ${\cal O}(\e)$ term of the 2-loop
sunset integral can be found in \cite{ints4loop}.
By an amusing relation specific to 3d,
namely the fact that the leading term of the 3d 1-loop scalar 2-point
integral is an 
arctan, whose derivative with respect to a mass looks like a
propagator with double mass, 
it is furthermore possible to 
relate the leading term of one of the 4-loop master integrals 
to a 3-loop case \cite{mikkoTrick}:
\ba
\ld\topoIVxtraNew\right|_{\rm const}&=&
\fr12\ld\threeMspecial\;\;\;\right|_{\rm const}\;.
\ea

There are however 2 master integrals (out of the 7 which get 
multiplied by a $1/\e$ from the prefactor) whose constant
term we do not yet know analytically. Let us denote their
leading parts by $x_2$ and $x_3$ (by naive power-counting, 
it is easy to see that both are UV finite),
\ba
\topoII = x_2+{\cal O}(\e) &,& 
\topoIIxtra = x_3+{\cal O}(\e)
\ea

Filling in the known expansions for the master integrals as well as
expanding the prefactors, higher poles 
cancel in the sum of diagrams, and we are left with a single
pole only:
\ba
d_A C_A^3 \fr{g^6}{(4\pi)^4} \(\fr{\bar\mu}{2m}\)^{8\e} 
\( \fr{p(\xi)}\e +{\cal O}(\e^0)\) \;.
\ea
\newcommand{\dilog}{{{\rm Li}_2}}
The polynomial $p$ is of order 6 in the gauge parameter $\xi$ 
and contains, besides a collection 
of numbers like $\pi^2$, $\ln2$ and dilogarithms, the two unknowns
$x_2$ and $x_3$. 
Clearly, in order for the result to be gauge independent,
all $\xi$-dependence has to vanish once $x_2$ and $x_3$ are known.
We can now reverse the argument and try to fix these constants by requiring
gauge independence. 
Inspecting the polynomial, it turns out to have a very simple
structure: 
\ba
p(\xi) &=& \alpha_G + (x_2-6x_3-b)\sum_{i=0}^6 c_i\,\xi^i \,\\
\label{eq:result}
\alpha_G &=& \fr{43}{96} - \fr{157}{6144}\,\pi^2 
\;\approx\; 0.195715\dots\;, 
\ea
where the $c_i$ are pure numbers and 
$b=\dilog\fr14+\dilog\fr15-3\dilog\fr25
+2(\ln2)^2-\fr32(\ln3)^2-(\ln5)^2
-2\ln2\ln5+3\ln3\ln5+\fr{\pi^2}8$. 
We have checked by numerical integration that
\ba \la{eq:zerocheck}
x_2-6x_3 &=& b \:\approx\;-0.00200966335\dots
\ea
to nine significant digits, hence establishing \eq\nr{eq:result}
as our main result for the logarithmic divergence of 3d pure 
gauge theory.


%

\vspace{4mm}
{\bf Acknowledgments}
I would like to thank K.~Kajantie, M.~Laine and A.~Vuorinen 
for numerous valuable discussions on the matter presented 
here, and KK and ML for an independent check of 
\eq\nr{eq:zerocheck}. 
This work was supported in parts by the DOE, 
under Cooperative Agreement no.~DF-FC02-94ER40818.



\end{document}